\pdfoutput=1

\documentclass[11pt]{article}

\usepackage[final]{acl}

\usepackage{times}
\usepackage{latexsym}
\usepackage{float}
\usepackage{amsmath}
\usepackage{array}
\usepackage{amssymb}
\usepackage{booktabs}
\usepackage{makecell}
\usepackage{tabularx}
\usepackage{subcaption}
\usepackage{booktabs}
\usepackage{siunitx}
\usepackage{colortbl}
\usepackage{xcolor}
\usepackage{caption} 
\usepackage[T1]{fontenc}

\usepackage[utf8]{inputenc}

\usepackage{microtype}

\usepackage{inconsolata}
\usepackage{multicol}
\usepackage{multirow}
\usepackage{graphicx}
\usepackage{parskip}
\usepackage{changepage}
\usepackage{afterpage}
\usepackage{xcolor}

\newcommand{\modified}[1]{{\color{black}{#1}}}

\usepackage[framemethod=tikz]{mdframed}
\newcounter{finding}
\newcommand{\finding}[1]{\refstepcounter{finding}
  \vspace{1.5mm}
 \begin{mdframed}[linecolor=gray,roundcorner=12pt,backgroundcolor=gray!15,linewidth=3pt,innerleftmargin=2pt, leftmargin=0cm,rightmargin=0cm,topline=false,bottomline=false,rightline = false]
 
  \textbf{Ans. to RQ\arabic{finding}:} #1
 \end{mdframed}
 \vspace{1.5mm}
}

\title{Personality-Guided Code Generation Using Large Language Models}

\author{Yaoqi Guo\textsuperscript{1}, Zhenpeng Chen\textsuperscript{2\textdagger}, Jie M.  Zhang\textsuperscript{3}, Yang Liu\textsuperscript{2}, Yun Ma\textsuperscript{1\textdagger} \\
        \textsuperscript{1}Peking University, China \\ \textsuperscript{2}Nanyang Technological University, Singapore \\ 
        \textsuperscript{3}King’s College London, United Kingdom \\
        \{ianwalls, mayun\}@pku.edu.cn, \\ \{zhenpeng.chen, yangliu\}@ntu.edu.sg, jie.zhang@kcl.ac.uk
        }

\begin{document}
\maketitle

\footnotetext{\textsuperscript{\textdagger}  Corresponding authors.}

\begin{abstract}
Code generation, the automatic creation of source code from natural language descriptions, has garnered significant attention due to its potential to streamline software development. Inspired by research that links task-personality alignment with improved development outcomes, we conduct an empirical study on personality-guided code generation using large language models (LLMs). Specifically, we investigate how emulating personality traits appropriate to the coding tasks affects LLM performance. We extensively evaluate this approach using seven widely adopted LLMs across four representative datasets. Our results show that personality guidance significantly enhances code generation accuracy, with improved pass rates in 23 out of 28 LLM-dataset combinations. Notably, in 11 cases, the improvement exceeds 5\%, and in 5 instances, it surpasses 10\%, with the highest gain reaching 12.9\%. Additionally, personality guidance can be easily integrated with other prompting strategies to further boost performance. 
We open-source our code and data at \url{https://github.com/IanWalls/Persona-Code}.
\end{abstract}

\section{Introduction}
Code generation, which aims to automatically produce source code from natural language descriptions, has attracted significant attention from both academia and industry due to its potential to streamline software development \citep{LLM4Code2}. The emergence of large language models (LLMs) has advanced this field by enabling the effective generation of complete, executable code \citep{chen2021evaluating, karmakar2021pre}. Additionally, specialized LLMs, such as CodeLlama \cite{codellama} and DeepSeek-Coder \cite{deepseekCoderV2}, have further refined these capabilities by focusing specifically on programming tasks. 

Previous research has observed that software development outcomes improve when individuals are assigned tasks that match their personality types \citep{capretz2015influence}. Furthermore, personality diversity within teams has been shown to correlate with higher-quality software deliverables \citep{ictasPieterseLE18, capretz2010we}.

In the code generation literature, LLMs are frequently tasked with role-playing as programmers to generate code \citep{LLM4Code2}. However, it is still unclear whether assigning these ``programmers'' with appropriate personalities and increasing personality diversity across tasks could further enhance code generation accuracy.

To fill this knowledge gap, we present an empirical study on personality-guided code generation using LLMs. Specifically, we first use GPT-4o, an advanced general-purpose LLM, to generate a programmer personality tailored to each coding task. Next, we assign various LLMs to emulate the roles of programmers with these generated personalities and evaluate whether this enhances their code generation accuracy.

We conduct a comprehensive evaluation of personality-guided code generation using seven widely-used LLMs and four well-recognized datasets. These LLMs, developed by leading vendors such as OpenAI, Meta, Alibaba, and DeepSeek, are extensively employed for code generation in both research and real-world applications \citep{LLM4SE1, LLM4SE2, GPT4omini, Codestral}. For personality characterization, we use the Myers-Briggs Type Indicator (MBTI) framework \citep{myers2003mbti}, which is widely applied in project management to align tasks with individual personality types and improve team dynamics \citep{capretz2010we}.

Our results demonstrate that personality guidance significantly enhances code generation accuracy, with pass rates improving in 23 out of 28 LLM-dataset combinations. In 11 cases, the improvement exceeds 5\%, and in 5 instances, it surpasses 10\%, with the highest gain reaching 12.9\%. 

Additionally, several factors appear to influence the effectiveness of personality-guided code generation, including LLM performance, dataset difficulty, and personality diversity. Specifically, moderate-performance LLMs benefit more from personality guidance compared to very strong or very weak models. Similarly, larger improvements are observed on datasets of moderate difficulty, as opposed to very easy or very difficult ones. Furthermore, greater personality diversity enhances the effectiveness of this approach, aligning with previous findings that diverse personality profiles in development teams are associated with higher-quality software outcomes \citep{ictasPieterseLE18, capretz2010we}.

Moreover, personality-guided code generation can be easily integrated with other prompting strategies. For instance, when combined with Chain of Thought \citep{CoT}, a widely-used prompting strategy in the code generation literature \citep{LLM4Code2}, we observe additional improvements in accuracy, with the highest gain reaching 13.8\%.

\section{Related Work}
This section summarizes existing work highly relevant to this paper.

\subsection{LLM for Code Generation}
The emergence of LLMs such as ChatGPT has profoundly transformed the landscape of automated code generation, making LLM-driven code generation a highly active area in both industry and the Natural Language Processing and Software Engineering communities \cite{LLM4Code2}. On one hand, researchers have explored the effectiveness of general-purpose LLMs such as ChatGPT \citep{ChatGPT}, GPT-4 \citep{GPT4}, and Llama \citep{llama3} for code generation. On the other hand, industry vendors have developed specialized LLMs designed to optimize code-related tasks, such as DeepSeek-Coder \citep{deepseekCoderV2} and CodeLlama \citep{codellama}. In this paper, to comprehensively evaluate the effectiveness of personality-guided code generation, we include both widely-used general-purpose LLMs and code-specific LLMs in our evaluation.

\begin{figure*}[!h]
    \centering
    \includegraphics[width=0.85\textwidth]{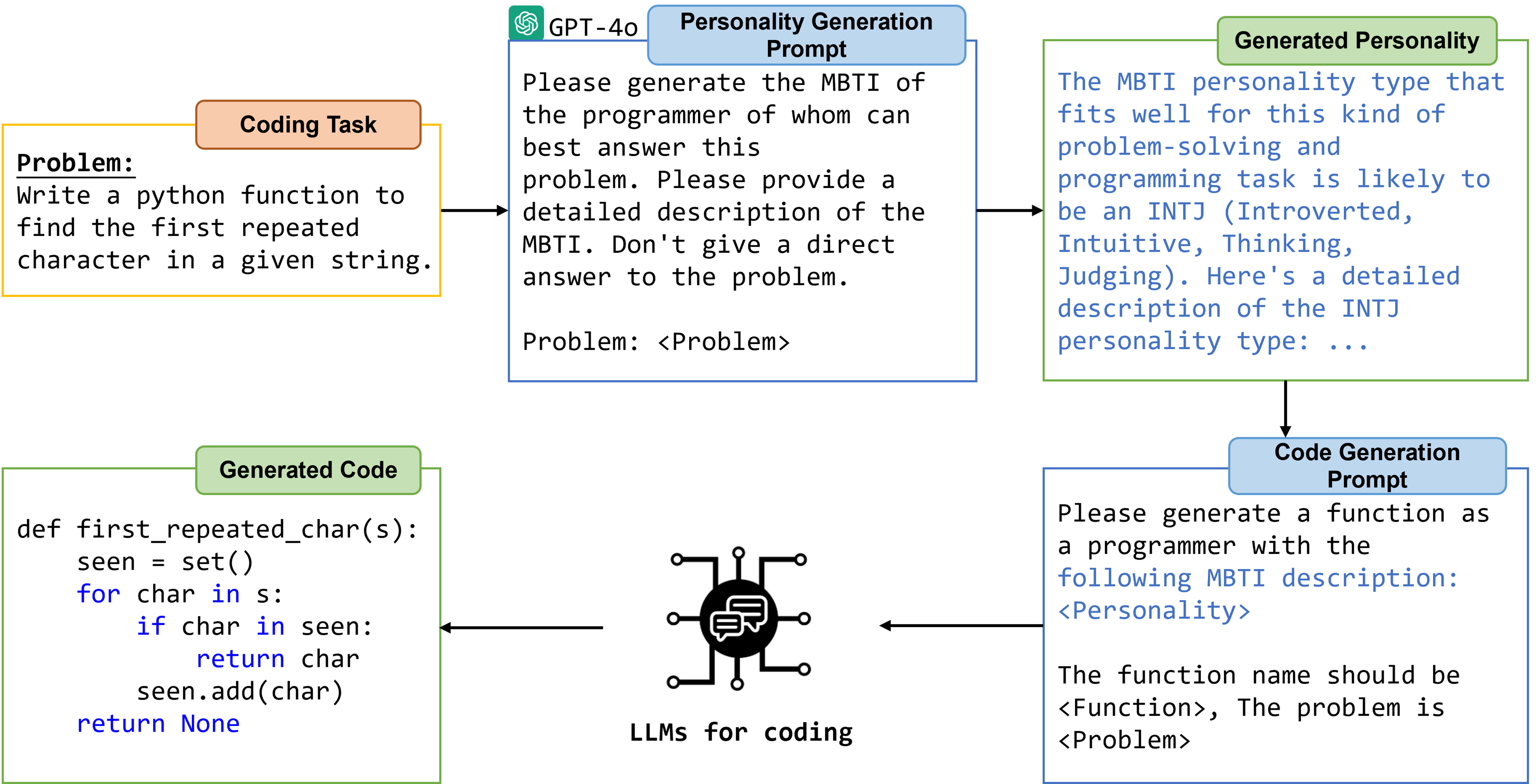} 
    \caption{Workflow of personality-guided code generation}
    \label{fig:prompt}
\end{figure*}

\subsection{Prompt Engineering for Code Generation}
Prompt engineering is an important strategy for improving the performance of LLM-based code generation \citep{LLM4SE2, huang2024effilearner}. It is widely recognized that LLMs can role-play to enhance performance in real-world tasks \citep{roleplay1, roleplay2, roleplay3}, and in the code generation literature, LLMs are often prompted to act as programmers when generating code \citep{LLM4Code2, liu2024large}. In addition, common prompting strategies such as few-shot learning \citep{LLM4Code1} and Chain of Thought \citep{CoT} are also widely adopted in code generation. In this paper, we extend the common practice of asking LLMs to act as programmers by equipping them with personalities tailored to the specific coding task. Additionally, we explore the integration of personality-guided code generation with established prompting strategies to further enhance performance.

\subsection{Personality of LLM}
Several studies have investigated the personality traits exhibited by LLMs \citep{psycho1, psycho2, psycho3, huang2023chatgpt}. For example, Huang et al. \citep{huang2023chatgpt} used trait theory, a psychological framework, to analyze the behavioral patterns of LLMs, finding that ChatGPT consistently exhibits an ENFJ personality. Building on the insight from prior research that diverse personality profiles within development teams are linked to higher-quality software outcomes \citep{ictasPieterseLE18, capretz2010we}, we explore the impact of assigning diverse personalities to LLMs when they tackle different coding tasks, aiming to ensure personality diversity and potentially enhance code generation accuracy.

\section{Personality-Guided Code Generation}\label{methodology}
This section presents our LLM-based pipeline for personality-guided code generation. As illustrated in Figure \ref{fig:prompt}, the pipeline consists of two key components: personality generation and code generation. The personality generation component is responsible for creating a programmer's personality suitable for addressing a given coding task. The code generation component then uses this generated personality to produce the code for the task. In the following, we define the coding task and provide a detailed description of each component.

\textbf{Task Definition:} We focus on function-level code generation tasks, which are among the most widely studied in the literature \citep{LLM4Code1}. These tasks typically provide a task description and require the generation of function code to solve the given problem. For each task, a test suite is provided, containing a collection of test cases. Each test case consists of a test input and its expected output. The generated code is considered to ``pass'' if it successfully passes the entire test suite, meaning it produces the correct output for all test cases.

\textbf{Personality Generation:} For a given coding task, we prompt GPT-4o, an advanced LLM for general text understanding, to generate an appropriate personality type most suited for solving the task. We adopt the Myers-Briggs Type Indicator (MBTI), a widely recognized personality framework that classifies individuals into 16 distinct types based on four dichotomies: Extraversion/Introversion, Sensing/Intuition, Thinking/Feeling, and Judging/Perceiving \citep{myers2003mbti}.
MBTI is popular for its accessibility and ease of interpretation, often used in project management to align tasks with individual personality types and enhance team dynamics \citep{capretz2010we}. In our approach, GPT-4o generates both the MBTI type and a detailed description tailored to the coding task.

\textbf{Code Generation:} Once the corresponding MBTI personality of a given task is generated, we move on to code generation. Our approach is applicable to various LLMs capable of generating coding. We prompt the LLM to take on the role of a programmer with the specified MBTI personality, providing a detailed description of that personality, and then it generates the code to solve the task. As previously mentioned, the generated code is considered a ``pass'' if it successfully passes all the test cases for the given coding task.

\section{Experimental Setup}
This section outlines the experimental setup used to evaluate personality-guided code generation.

\subsection{Research Questions (RQs)}
We aim to answer the following five RQs to evaluate personality-guided code generation.

\noindent \textbf{RQ1 (Effectiveness):} How effective is personality-guided code generation in enhancing generation accuracy?

\noindent \textbf{RQ2 (Influencing factors):} What potential factors influence the effectiveness of personality-guided code generation?

\noindent \textbf{RQ3 (Combination with other strategies):} Can personality-guided code generation be combined with other prompting strategies to further improve generation accuracy?

\noindent \textbf{RQ4 (Prompt design):} How does including the detailed personality description during code generation affect the accuracy of the generated code?

\modified{

\noindent \textbf{RQ5 (Personality modeling):} How do different personality modeling methods impact the effectiveness of our approach?

}

\subsection{Datasets}
We evaluate personality-guided code generation using four widely recognized datasets: MBPP Sanitized \citep{MBPP}, MBPP+ \citep{EvalPlus}, HumanEval+ \citep{EvalPlus}, and APPS \citep{APPS}. In the following, we provide a brief description of each dataset.

\noindent $\bullet$ \emph{MBPP Sanitized} includes 427 crowd-sourced Python problems designed for entry-level programmers, each with a task description, code solution, and several automated test cases. 

\noindent $\bullet$ \emph{MBPP+} improves upon MBPP by fixing ill-formed problems and incorrect implementations, while expanding the test suite by 35 times for more robust evaluation. 

\noindent $\bullet$ \emph{HumanEval+} offers 164 manually curated Python problems, each featuring a function signature, docstring, code body, and multiple unit tests to detect errors that LLMs might miss. 

\noindent $\bullet$ \emph{APPS} presents a comprehensive benchmark of 10K Python problems across varying difficulty levels. Given the large size of the dataset, we randomly sample 500 problems from the interview-level set to balance evaluation depth with computational efficiency.

\subsection{LLMs Used for Code Generation}
We adopt seven LLMs for evaluation, consisting of four general-purpose LLMs and three specifically designed for code-related tasks. The general LLMs include GPT-4o \cite{GPT4o}, GPT-4o mini \cite{GPT4omini}, Llama3.1 \cite{llama3}, and Qwen-Long \cite{Qwen2}, while the code-specific LLMs include DeepSeek-Coder \cite{deepseekCoderV2}, Codestral \cite{Codestral}, and CodeLlama \cite{codellama}. 

Table \ref{tab:models} lists the information of these LLMs. These LLMs are developed by leading vendors such as OpenAI, Meta, Alibaba, and DeepSeek. All of them are widely used for code generation in research and practical applications \citep{LLM4SE1, LLM4SE2, GPT4omini, Codestral}. 

\subsection{Evaluation Metric}
We evaluate code generation accuracy by calculating the pass rate across all tasks in the dataset. An LLM is considered to pass a coding task if the code it generates successfully passes all test cases for that task. Considering the non-determinism of LLMs in code generation\citep{ouyang2025empirical}, to ensure the reliability of our results, we run each LLM on each dataset three times and report the average pass rate as the final outcome.
Specifically, the pass rate $P$ of an LLM on a dataset is calculated as:
$$P=\frac{1}{3} \sum_{i=1}^{3} \frac{c_i}{cnt},$$
where $cnt$ represents the total number of tasks in the dataset, and $c_i$ is the number of tasks successfully passed by the LLM in the $i$th run.

\begin{table}
\footnotesize
\centering
\begin{tabular}{lrrr}
\hline
\textbf{LLM} & \textbf{Size} & \textbf{Institution} & \textbf{Date} \\
\hline
GPT-4o & - & OpenAI & 2024-05 \\
GPT-4o mini & - & OpenAI & 2024-07 \\
Llama3.1 & 70B & Meta & 2024-07 \\
Qwen-Long & - & Alibaba & 2024-05 \\
DeepSeek-Coder V2 & 236B & DeepSeek & 2024-06 \\
Codestral & 22B & Mistral AI & 2024-05 \\
CodeLlama & 13B & Meta & 2023-08 \\
\hline
\end{tabular}
\caption{LLMs used for code generation}
\label{tab:models}
\end{table}

\section{Results}
This section answers our research questions with the experimental results.

\begin{table*}[h]
\centering
\small
\begin{tabularx}{\textwidth}{l|l*{4}{>{\centering\arraybackslash}X}c}
\toprule
 \multicolumn{1}{c|}{LLM}& \multicolumn{1}{c}{} & \makecell[c]{MBPP\\Sanitized} & \multicolumn{1}{c}{MBPP+} & \multicolumn{1}{c}{HumanEval+} & \multicolumn{1}{c}{APPS} & \multicolumn{1}{c}{\makecell[c]{Average \\ Change}} \\
\midrule
\multirow{3}{*}{GPT-4o} & Direct  & 78.2\% & 71.2\% & 84.8\% & 46.2\% & \multirow{3}{*}{\centering{\textbf{↑ 1.2\%}}} \\
 & MBTI  & 84.3\% & 72.7\% & 82.9\% & 45.2\% & \\
 & Change & \cellcolor{gray!60}{\textbf{↑ 6.1\%}} & \cellcolor{gray!60}{\textbf{↑ 1.5\%}} & \cellcolor{gray!20}{\textbf{↓ 1.9\%}} & \cellcolor{gray!20}{\textbf{↓ 1.0\%}} & \\
\midrule
\multirow{3}{*}{GPT-4o mini} & Direct  & 69.3\% & 69.4\% & 80.5\% & 34.6\% & \multirow{3}{*}{\centering{\textbf{↑ 4.9\%}}} \\
 & MBTI  & 82.2\% & 71.7\% & 82.3\% & 37.2\% & \\
 & Change & \cellcolor{gray!60}{\textbf{↑ 12.9\%}} & \cellcolor{gray!60}{\textbf{↑ 2.3\%}} & \cellcolor{gray!60}{\textbf{↑ 1.8\%}} & \cellcolor{gray!60}{\textbf{↑ 2.6\%}} & \\
\midrule
\multirow{3}{*}{Llama3.1} & Direct  & 69.8\% & 66.7\% & 72.0\% & 18.4\% & \multirow{3}{*}{\centering{\textbf{↑ 5.3\%}}} \\
 & MBTI  & 81.0\% & 69.2\% & 72.6\% & 25.2\% & \\
 & Change & \cellcolor{gray!60}{\textbf{↑ 11.2\%}} & \cellcolor{gray!60}{\textbf{↑ 2.5\%}} & \cellcolor{gray!60}{\textbf{↑ 0.6\%}} & \cellcolor{gray!60}{\textbf{↑ 6.8\%}} & \\
\midrule
\multirow{3}{*}{Qwen-Long} & Direct  & 68.4\% & 67.7\% & 76.8\% & 10.2\% & \multirow{3}{*}{\centering{\textbf{↑ 6.5\%}}} \\
 & MBTI  & 80.8\% & 71.2\% & 78.7\% & 18.2\% & \\
 & Change & \cellcolor{gray!60}{\textbf{↑ 12.4\%}} & \cellcolor{gray!60}{\textbf{↑ 3.5\%}} & \cellcolor{gray!60}{\textbf{↑ 1.9\%}} & \cellcolor{gray!60}{\textbf{↑ 8.0\%}} & \\
\midrule
\multirow{3}{*}{DeepSeek-Coder V2} & Direct  & 74.9\% & 71.4\% & 80.5\% & 39.4\% & \multirow{3}{*}{\centering{\textbf{↑ 0.6\%}}} \\
 & MBTI  & 85.7\% & 72.2\% & 76.2\% & 34.4\% & \\
 & Change & \cellcolor{gray!60}{\textbf{↑ 10.8\%}} & \cellcolor{gray!60}{\textbf{↑ 0.8\%}} & \cellcolor{gray!20}{\textbf{↓ 4.3\%}} & \cellcolor{gray!20}{\textbf{↓ 5.0\%}} & \\
\midrule
\multirow{3}{*}{Codestral} & Direct  & 64.2\% & 61.2\% & 75.6\% & 15.8\% & \multirow{3}{*}{\centering{\textbf{↑ 5.3\%}}} \\
 & MBTI  & 73.8\% & 64.9\% & 76.8\% & 22.6\% & \\
 & Change & \cellcolor{gray!60}{\textbf{↑ 9.6\%}} & \cellcolor{gray!60}{\textbf{↑ 3.7\%}} & \cellcolor{gray!60}{\textbf{↑ 1.2\%}} & \cellcolor{gray!60}{\textbf{↑ 6.8\%}} & \\
\midrule
\multirow{3}{*}{CodeLlama} & Direct  & 43.3\% & 42.4\% & 32.9\% & 1.4\% & \multirow{3}{*}{\centering{\textbf{↑ 3.9\%}}} \\
 & MBTI  & 46.8\% & 52.4\% & 29.9\% & 6.4\% & \\
 & Change & \cellcolor{gray!60}{\textbf{↑ 3.5\%}} & \cellcolor{gray!60}{\textbf{↑ 10.0\%}} & \cellcolor{gray!20}{\textbf{↓ 3.0\%}} & \cellcolor{gray!60}{\textbf{↑ 5.0\%}} & \\
\bottomrule
\end{tabularx}
\caption{Comparison of pass rates for LLMs with and without personality guidance across different datasets}
\label{tab:performance}
\end{table*}

\subsection{RQ1: Effectiveness}
Table \ref{tab:performance} presents the comparison of pass rate for LLMs with and without personality guidance across different datasets. Specifically, for each LLM on each dataset, we evaluate two approaches: directly prompting the LLM to generate code as a programmer (the ``Direct'' row) and using the personality-guided method (the ``MBTI'' row). We report the pass rates for both approaches on each dataset, along with the change in performance introduced by personality guidance.

Overall, the personality-guided approach improves the pass rate of code generation in 23 out of 28 combinations of LLMs and datasets. In 11 combinations, the improvement exceeds 5\%, and in 5 combinations, it surpasses 10\%. Notably, the pass rate of GPT-4o mini on the MBPP Sanitized dataset increases by 12.9\%.

Additionally, we observe that personality-guided code generation enhances the pass rate for all LLMs considered, with average improvements ranging from 0.6\% to 6.5\% across all datasets. Specifically, the average pass rates of Llama3.1, Qwen-Long, and Codestral improve by 5.3\%, 6.5\%, and 5.3\%, respectively.

\finding{Personality-guided code generation significantly enhances generation accuracy, improving pass rates in 23 out of 28 LLM-dataset combinations. In 11 cases, the improvement exceeds 5\%, and in 5 cases, it surpasses 10\%, with GPT-4o mini showing a 12.9\% gain on the MBPP Sanitized dataset. }

\subsection{RQ2: Influencing Factors}

\begin{figure*}[ht]
  \centering
  \begin{subfigure}[b]{0.24\linewidth}
    \includegraphics[width=\linewidth]{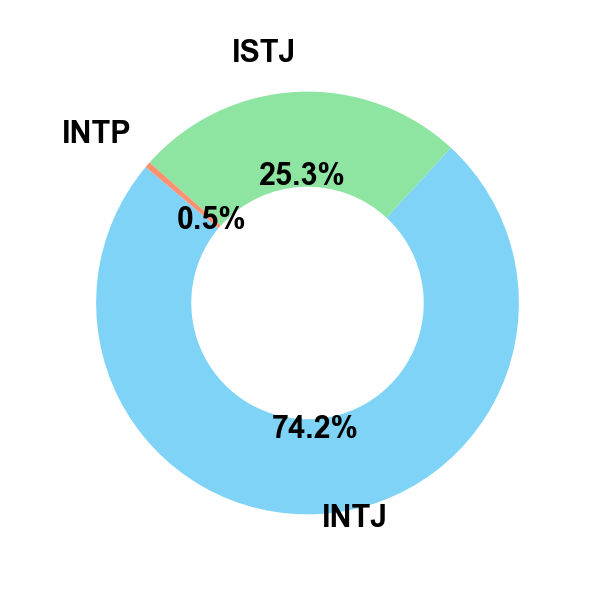}
    \caption{MBPP Sanitized}
  \end{subfigure} \hfill
  \begin{subfigure}[b]{0.24\linewidth}
    \includegraphics[width=\linewidth]{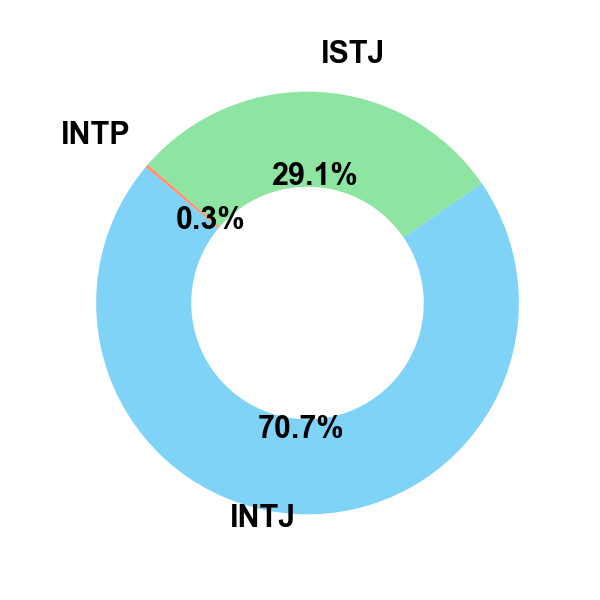}
    \caption{MBPP+}
  \end{subfigure} \hfill
  \begin{subfigure}[b]{0.24\linewidth}
    \includegraphics[width=\linewidth]{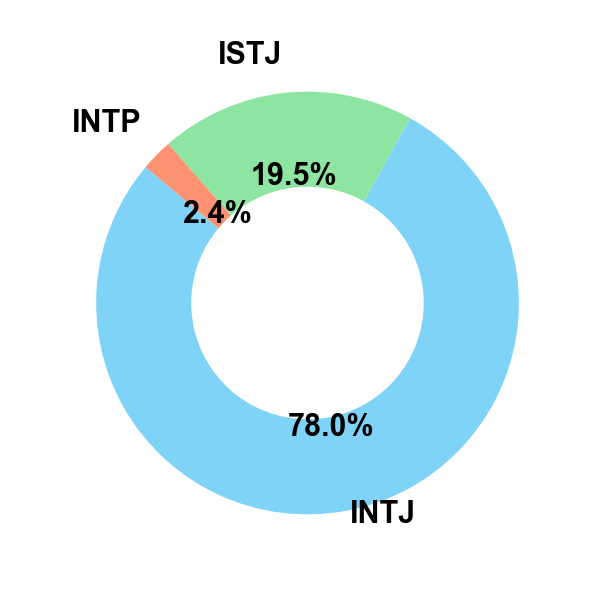}
    \caption{HumanEval+}
  \end{subfigure} \hfill
  \begin{subfigure}[b]{0.24\linewidth}
    \includegraphics[width=\linewidth]{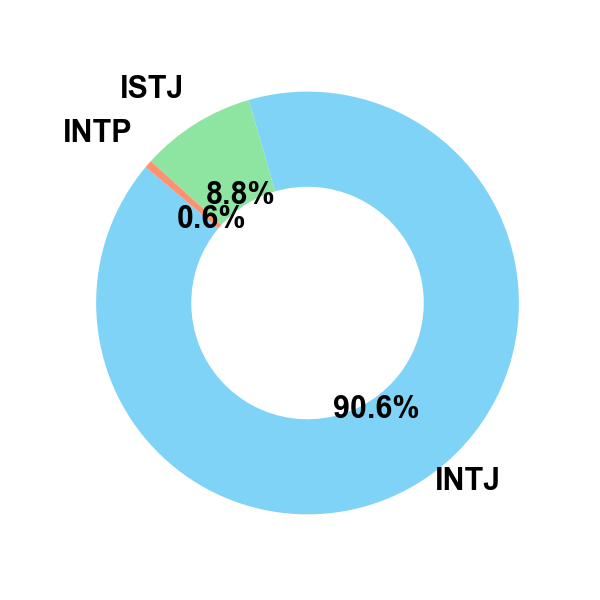}
    \caption{APPS}
  \end{subfigure}
  \caption{Distribution of MBTI types generated by GPT-4o for each dataset}
  \label{fig:Distribution}
\end{figure*}

Furthermore, we explore the potential factors influencing the effectiveness of personality-guided code generation, examining them from both the model and dataset perspectives.

\subsubsection{Model Perspective Analysis}
Table \ref{tab:performance} shows that only GPT-4o, DeepSeek-Coder V2, and CodeLlama exhibit slight decreases in pass rates in one or two cases. Additionally, in the ``Direct'' mode, GPT-4o and DeepSeek-Coder consistently achieve the highest pass rates among all the LLMs, while CodeLlama has the lowest.

Based on these observations, we conclude that LLMs with moderate baseline performance may benefit more from personality guidance compared to those with either very strong or very weak baseline performance. This is reasonable, as models with excellent performance may already be near their potential, while those with low performance may require more fundamental improvements beyond personality guidance.

\begin{table}[t]
\centering
\small
\begin{tabular}{l|cccc}
\toprule
\textbf{Uniform MBTI} & ENFJ & ENFP & ENTJ & ENTP \\
\textbf{Pass rate} & 67.9\% & 67.9\% & 66.7\% & 65.7\% \\
\midrule
\textbf{Uniform MBTI} & ESFJ & ESFP & ESTJ & ESTP \\
\textbf{Pass rate} & 67.9\% & 67.4\% & 67.9\% & 67.6\% \\
\midrule
\textbf{Uniform MBTI} & INFJ & INFP & INTJ & INTP \\
\textbf{Pass rate} & 67.4\% & 67.4\% & 66.6\% & 68.2\% \\
\midrule
\textbf{Uniform MBTI} & ISFJ & ISFP & ISTJ & ISTP \\
\textbf{Pass rate} & 67.1\% & 68.4\% & 66.6\% & 67.4\% \\
\midrule
\multicolumn{5}{c}{\textbf{Direct: 68.4\% Diverse MBTI: 80.8\%}} \\
\bottomrule
\end{tabular}
\caption{Pass rates for Qwen-Long on the MBPP Sanitized dataset when using uniform MBTI personalities, compared to direct code generation (Direct) and our diverse personality-guided approach (Diverse MBTI)}
\label{tab:DifMBTI}
\end{table}

\subsubsection{Dataset Perspective Analysis}
From Table \ref{tab:performance}, we observe that pass rate decreases occur only in the HumanEval+ and APPS datasets. Notably, HumanEval+ is the easiest dataset, as five LLMs achieve a pass rate higher than 75\%. In contrast, APPS is the most challenging, with no LLM achieving a pass rate above 40\%.

These observations suggest that the difficulty level of the dataset influences the effectiveness of personality-guided code generation. It is reasonable that on easier tasks, where models already perform well (as seen with HumanEval+), personality guidance may offer limited improvement, or in some cases, a slight decrease. On highly challenging tasks like APPS, where baseline performance is lower, there may be more room for improvement, but the complexity of the task might limit the potential gains. Fortunately, the personality-guided approach achieves more than a 5\% improvement in pass rates for four out of seven LLMs.

Furthermore, we analyze the diversity of personality distributions across each dataset. Figure \ref{fig:Distribution} presents the MBTI personality types assigned by GPT-4o for each dataset. We observe that HumanEval+ and APPS exhibit the least diversity, with 78.0\% and 90.6\% of tasks assigned the INTJ personality, respectively. This suggests that personality diversity may be a potential factor influencing the effectiveness of personality-guided code generation. The more diverse the assigned personalities, the more effective this approach tends to be.

To further demonstrate the impact of personality diversity, we set up an additional experiment to investigate the effect of assigning a uniform MBTI personality to all tasks. We select Qwen-Long as the test model because it exhibits the highest average improvement from personality guidance. Additionally, we use the MBPP Sanitized dataset, where Qwen-Long shows the greatest improvement. The significant improvement on this dataset provides a clear baseline, allowing us to better observe the impact of reducing personality diversity.

Since the MBTI framework includes 16 personality types, we consider 16 uniform personality approaches. For each approach, Qwen-Long is prompted to take on the role of a uniform MBTI personality across all coding tasks. Table \ref{tab:DifMBTI} presents the results. The code generation accuracy varied slightly across the uniform personalities, with the highest pass rate for ISFP (68.4\%) and the lowest for ENTP (65.7\%), both close to the direct prompting rate (68.4\%). In contrast, our diverse personality-guided approach, which assigns suitable MBTI types for each task, achieves a significantly higher pass rate of 80.8\%. 

\begin{figure}[!h]
    \centering
    \includegraphics[width=0.3\textwidth]{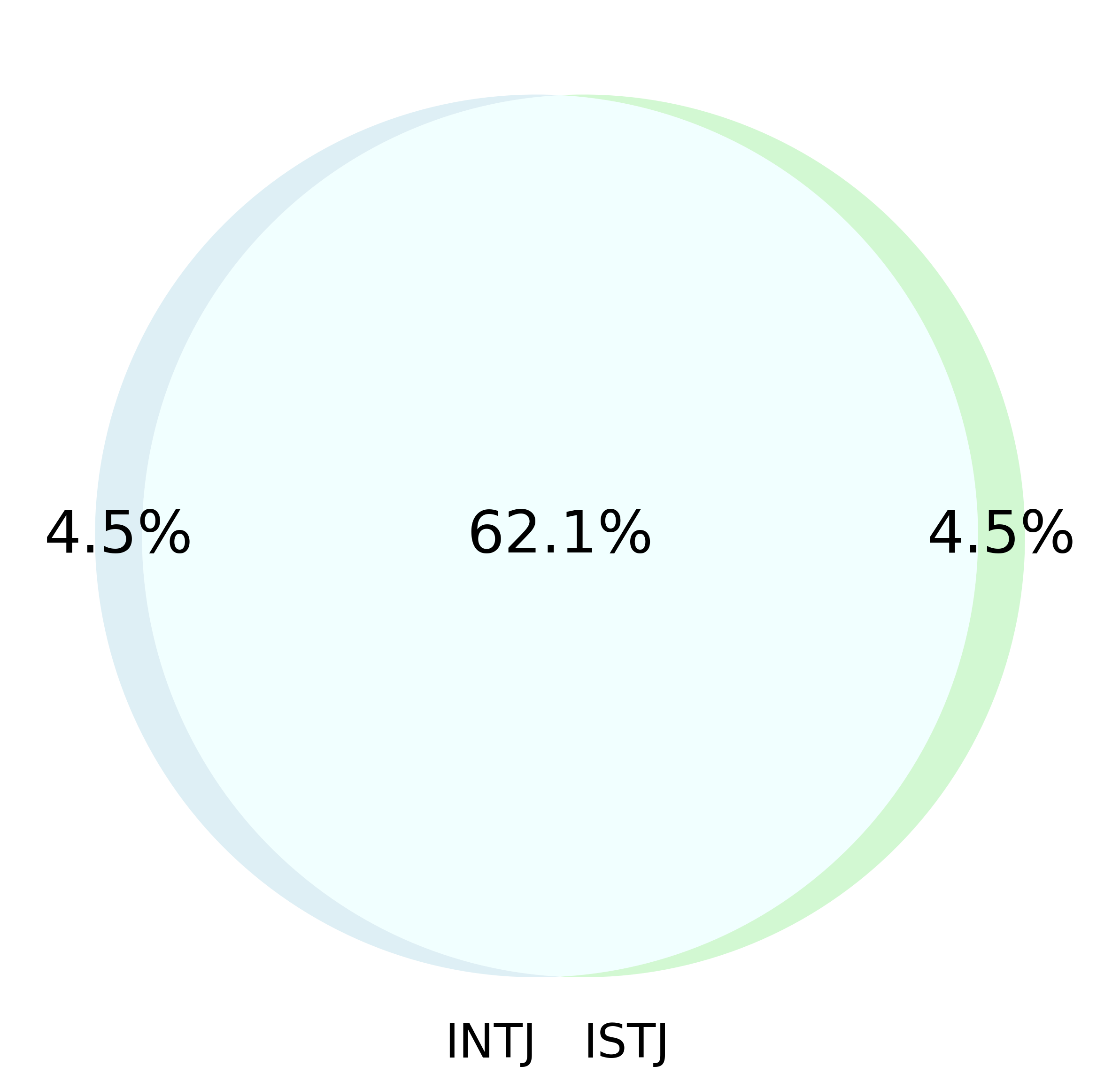} 
    \caption{Venn diagram illustrating the tasks solved by INTJ versus ISTJ types}
    \label{fig:Venn}
\end{figure}

\modified{Moreover, while INTJ dominates in Figure 2, personality diversity remains essential. To assess the impact of different MBTI types on code generation performance, we create a Venn diagram at Figure \ref{fig:Venn} illustrating the tasks solved by INTJ versus ISTJ types. A closer analysis of the data in Table \ref{tab:DifMBTI} reveals that 62.1\% of tasks can be solved by either INTJ or ISTJ, while 4.5\% are uniquely solvable by INTJ and another 4.5\% by ISTJ. This highlights the complementary strengths of different personality types and the importance of diversity in addressing a broader range of tasks.}

\finding{The LLM's performance, dataset difficulty, and personality diversity are potential factors influencing the effectiveness of personality-guided code generation. LLMs with moderate performance benefit more than those with either very strong or very weak performance. Similarly, improvements are greater on datasets of moderate difficulty compared to those that are very easy or very difficult. Additionally, greater personality diversity tends to enhance the effectiveness of personality-guided code generation.}

\begin{table*}[h]
\centering
\small
\begin{tabularx}{\textwidth}{l|*{5}{>{\centering\arraybackslash}X}}
\toprule
\textbf{LLM} & \textbf{Personality} & \textbf{Three-shot} & \textbf{Three-shot +Personality} & \textbf{CoT} & \textbf{CoT+ Personality} \\ 
\midrule
GPT-4o & 84.3\% & 77.3\% & 86.7\% & 77.0\% & 85.7\% \\ 
GPT-4o mini & 82.2\% & 69.6\% & 82.2\% & 71.9\% & 81.5\% \\ 
Llama3.1 & 74.9\% & 67.7\% & 78.9\% & 67.4\% & 79.9\% \\ 
Qwen-Long & 80.8\% & 66.7\% & 76.6\% & 69.6\% & 82.2\% \\ 
DeepSeek-Coder V2 & 85.7\% & 76.1\% & 83.8\% & 72.4\% & 83.8\% \\ 
Coderstral & 73.8\% & 65.3\% & 72.8\% & 66.0\% & 74.5\% \\ 
CodeLlama & 46.8\% & 46.6\% & 47.1\% & 30.4\% & 49.6\% \\ 
\bottomrule
\end{tabularx}
\caption{Comparison of pass rates achieved by our approach, existing prompting strategies, and their combination}
\label{tab:CoTFewShot}
\end{table*}

\subsection{RQ3: Combination with Other Strategies}
This RQ investigates whether the effectiveness of personality-guided code generation can be further enhanced by combining it with existing prompting strategies. Specifically, we consider two widely used techniques: few-shot learning and Chain of Thought (CoT), both of which are popular methods for code generation \citep{LLM4Code2}. Few-shot learning provides examples to guide the model's response, helping it generalize from limited data, while CoT encourages the model to generate intermediate reasoning steps, improving complex problem-solving accuracy.

For few-shot learning, we use a three-shot approach, a widely adopted setting in the code generation literature \citep{LLM4Code1, shotNum}; for CoT, we prompt the LLM to think step by step \citep{CoT}. We select the MBPP Sanitized dataset for this experiment, as it shows the most significant improvement with our personality-guided approach, making it an ideal candidate to explore potential further gains.

Table \ref{tab:CoTFewShot} presents the results. First, we find that personality-guided approach (the ``Personality'' column) consistently outperforms both three-shot learning and CoT strategies across all seven LLMs.

Next, we compare our approach to its combination with existing strategies. Three-shot + Personality outperforms the personality-guided approach in only three of the seven LLMs, while CoT + Personality achieves better results in five of the seven LLMs, with improvements ranging from 3.8\% to 12.9\%. Thus, combining CoT and personality guidance is a a promising solution, yielding improvements of 6.3\% to 13.8\% over direct prompting, depending on the LLM used.

\finding{Personality-guided code generation consistently outperforms both three-shot learning and Chain of Thought strategies across all seven LLMs. Additionally, combining Chain of Thought with personality guidance further enhances code generation accuracy, with improvements ranging from 6.3\% to 13.8\%, depending on the LLM used.}

\subsection{RQ4: Prompt Design}

\begin{table}[t]
\centering
\small
\begin{tabular}{lccc}
\hline
\textbf{LLM} & \textbf{Direct} & \makecell[c]{\textbf{Full}\\ \textbf{Prompt}} & \makecell[c]{\textbf{Short}\\ \textbf{Prompt}} \\
\hline
GPT-4o & 78.2\% & 84.3\% & 82.9\% \\
GPT-4o mini & 69.3\% & 82.2\% & 80.1\% \\
Llama3.1 & 69.8\% & 81.0\% & 72.4\% \\
Qwen-Long & 68.4\% & 80.8\% & 73.3\% \\
DeepSeek-Coder & 74.9\% & 85.7\% & 83.1\% \\
Coderstral & 64.2\% & 73.8\% & 72.1\% \\
CodeLlama & 43.3\% & 46.8\% & 43.1\% \\
\bottomrule
\end{tabular}
\caption{Comparison of pass rates between Direct prompting, our approach using the full MBTI description (Full Prompt), and using only the MBTI type (Short Prompt)}
\label{tab:LongShort}
\end{table}

In Section \ref{methodology}, we describe that during the code generation process, we provide both the MBTI type and its detailed description. This is based on the assumption that a more detailed personality description may help the LLM better role-play, potentially improving performance. However, existing research suggests that longer prompts can negatively affect LLM performance on the same task \cite{LengthAffect}.

To address this, in this RQ, we evaluate whether providing a detailed description alongside the MBTI type yields better results than using a shorter prompt that only indicates the MBTI type (e.g., INTJ). As with RQ3, we select the MBPP Sanitized dataset for this experiment.

Table \ref{tab:LongShort} presents the results. We find that using the full MBTI description (the ``Full Prompt'' column) consistently achieves higher pass rates than using only the MBTI type (the ``Short Prompt'' column) across all seven LLMs. On average, the full MBTI description improves the pass rate by 3.94\%. 

Although using only the MBTI type underperforms compared to the full description, it still outperforms the direct prompt in six out of seven LLMs, highlighting the effectiveness of personality guidance.

\modified{Furthermore, we experiment with using a template of 16 general MBTI descriptions (generated by GPT-4) in our prompt, instead of having LLMs generate them each time for every coding task. Here, we use the MBPP Sanitized dataset and Qwen-Long, as it shows the highest average improvement from personality guidance. The results indicate that using the general template results in a pass rate of 65.5\%, significantly lower than the 80.8\% achieved by our default approach.}

\finding{Using the full MBTI description (i.e., the default setting in our approach) consistently outperforms using only the MBTI type across all seven LLMs, with an average performance improvement of 3.94\%.}

\begin{table}[t]
\centering
\small
\begin{tabular}{lcc}
\hline
\textbf{LLM} & \makecell[c]{\textbf{MBTI}} & \makecell[c]{\textbf{Big Five}} \\
\hline
GPT-4o & 84.3\% & 82.9\% \\
GPT-4o mini & 82.2\% & 69.0\% \\
Llama3.1 & 81.0\% & 72.1\% \\
Qwen-Long & 80.8\% & 71.4\% \\
DeepSeek-Coder & 85.7\% & 72.8\% \\
Coderstral & 73.8\% & 63.4\% \\
CodeLlama & 46.8\% & 42.4\% \\
\bottomrule
\end{tabular}
\caption{Comparison of pass rates between using MBTI and Big Five Personality}
\label{tab:BigFive}
\end{table}

\subsection{RQ5: Personality Modeling}
\modified{
This RQ examines how the choice of personality modeling methods influences the effectiveness of our approach. To this end, we compare our default approach (i.e., using MBTI) with the Big Five Personality model~\citep{BigFive}, another widely used personality modeling framework. The Big Five Personality model includes five dimensions: openness, conscientiousness, extroversion, agreeableness, and neuroticism.

We use the same experimental setup and prompts as those used for MBTI prompting, with the only change being the personality model, which is switched to the Big Five Personality. As in RQ3 and RQ4, we select the MBPP Sanitized dataset for this experiment. The comparison results are shown in Table \ref{tab:BigFive}. Our findings indicate that MBTI-based personality prompting outperforms Big Five Personality prompting across all evaluated LLMs.

This result suggests that MBTI is more suitable for personality modeling in our framework. This can be attributed to the different perspectives emphasized by the two modeling methods. Among the five dimensions of the Big Five Personality model, only conscientiousness is strongly linked to coding performance. In contrast, the MBTI offers a structured framework for modeling cognitive preferences, making it particularly well-suited for code generation. Each MBTI dimension reflects distinct cognitive approaches: for example, Sensing emphasizes detail-oriented problem-solving for debugging, Intuition fosters abstract thinking for algorithm design, Thinking ensures logical precision, and Feeling takes values into account for user-centric decisions.

Code generation uniquely requires logical precision, abstract reasoning, and context-sensitive problem-solving—traits that align well with MBTI dimensions. Embedding these cognitive traits into LLMs allows for task-specific alignment, much like matching tasks to human cognitive strengths, providing a theoretical basis for the enhanced effectiveness of code generation.

\finding{MBTI is better suited for personality modeling in our framework than the Big Five Personality Model. Across all evaluated LLMs, using MBTI results in higher code generation accuracy compared to the Big Five.}

}

\section{Discussion}
\modified{
We further discuss the implications based on our findings:  (1) Our study demonstrates that introducing personality traits into LLMs significantly enhances code generation accuracy. It offers practical solutions for developers to improve LLMs for better code generation, and suggests that LLM performance is not purely computational but can be improved by mimicking human cognitive processes, transforming LLMs into more nuanced, context-aware problem solvers. (2) Our results emphasize the importance of personality diversity in improving task-specific performance, highlighting its potential to address a broader range of challenges effectively. (3) The observed synergy between personality guidance and strategies like Chain of Thought underscores its modularity, showing that personality guidance can integrate seamlessly with other techniques to enhance LLM reasoning and problem-solving.
}

\section{Conclusion}
This paper presents a large-scale empirical study on personality-guided code generation using LLMs. While existing research typically involves LLMs role-playing as programmers to generate code, this study investigates whether assigning these ``programmers'' with appropriate personalities can further improve code generation accuracy. To explore this, we conduct an extensive evaluation using four widely-adopted datasets and seven advanced LLMs developed by leading vendors. Our results show that personality guidance significantly boosts code generation accuracy, with pass rates improving in 23 out of 28 LLM-dataset combinations. Notably, in 11 cases, the improvement exceeds 5\%, and in 5 instances, it surpasses 10\%, with the highest gain reaching 12.9\%.

\section{Limitations}
As an empirical study, this paper has several limitations. 
First, the personality traits examined are limited to the MBTI framework. While MBTI is widely used, relying solely on it may not capture the full complexity of personality traits and their potential impact on LLM performance. 
Second, although we evaluated seven LLMs, including both general-purpose and code-task-specific models, the generalizability of our findings to other LLMs requires further investigation. 
Third, our study focuses on function-level code generation across four datasets, a common area in the literature. In future work, we plan to extend our evaluation to more complex code generation tasks to broaden the scope of our findings.
\modified{Fourth, since we do not have ground-truth personality labels for each task, we evaluate the predictive accuracy of GPT-4o's generated personalities indirectly through their impact on code generation accuracy. Thus we cannot explicitly calculate the theoretical upper bound score of personality guided code generation.}

\section*{Acknowledgments}
Jie M. Zhang is supported by the ITEA Genius and ITEA GreenCode projects, funded by InnovateUK. Zhenpeng Chen and Yang Liu are supported by the National Research Foundation Singapore and DSO National Laboratories under the AI Singapore Programme (AISG Award No. AISG2-RP-2020-019); by the National Research Foundation Singapore and the Cyber Security Agency of Singapore under the National Cybersecurity R\&D Programme (NCRP25-P04-TAICeN); and by the National Research Foundation, Prime Minister’s Office, Singapore under the Campus for Research Excellence and Technological Enterprise (CREATE) programme. Any opinions, findings, conclusions, or recommendations expressed in this paper are those of the authors and do not reflect the views of the National Research Foundation Singapore or the Cyber Security Agency of Singapore.
\bibliography{custom}

\appendix

\section{Prompt and Personality Example}

\begin{figure*}[!h]
    \centering
    \includegraphics[width=0.92\textwidth]{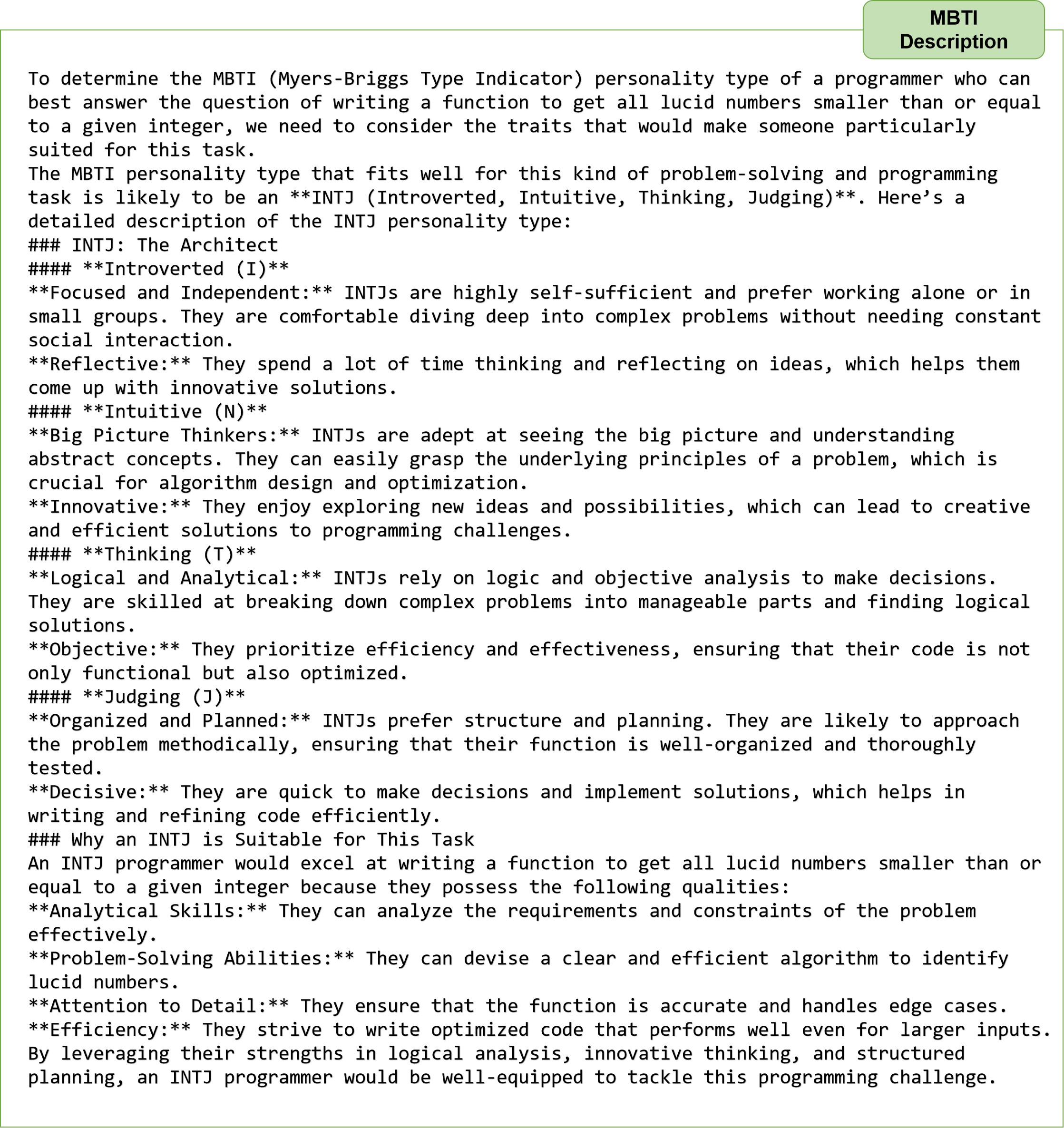} 
    \caption{Personality example recommended by GPT-4o}
    \label{fig:PersonalityPrompt}
\end{figure*}

\begin{figure*}[!h]
    \centering
    \includegraphics[width=0.92\textwidth]{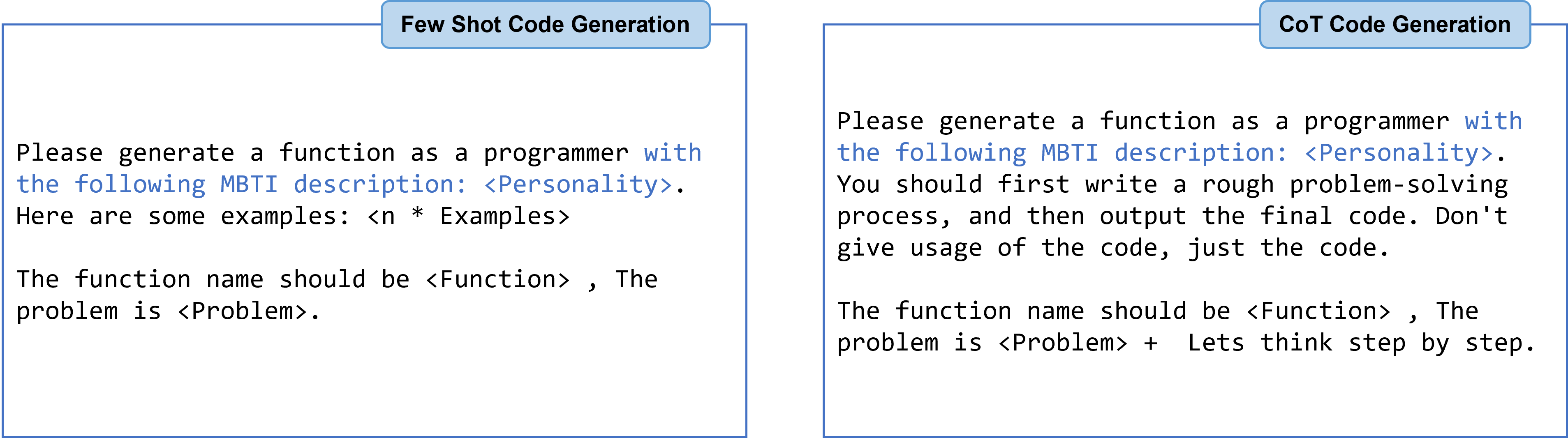} 
    \caption{Few-shot and CoT prompt structures}
    \label{fig:CoTFewshot}
\end{figure*}

\modified{
The prompts used in MBTI prompting are listed in Figure~\ref{fig:prompt}. The personality recommendation example is listed in Figure~\ref{fig:PersonalityPrompt}.
The few-shot prompting and CoT prompting structures are listed in Figure~\ref{fig:CoTFewshot}.
}

\section{Impact of Personality-Generation LLM}

\modified{

This section tries to answer the question that if we use other LLMs to generate personality, how would this impact the effectiveness of personality-guided code generation?

In Section \ref{methodology}, we describe that GPT-4o is used for personality generation. This RQ evaluates this setting by comparing it to a setting where the code-generation LLM is also used for personality generation. For example, when Qwen-Long is used for code generation, it also generates its own personality for each task. CodeLlama is excluded from the comparison because it lacks the capability to generate appropriate personality. As with RQ3 and RQ4, we use the MBPP Sanitized dataset for this experiment.

Table \ref{tab:SelfSota} shows the comparison results. We find that using GPT-4o consistently for personality generation (the ``GPT-4o'' column) outperforms using each LLM individually (the ``Self'' column) in four out of five LLMs. For GPT-4o mini, using it for both personality and code generation results in a slightly higher pass rate (0.5\%) than using GPT-4o for personality generation. For the remaining four LLMs, the average improvement of using GPT-4o over the code-generation LLM is 4.1\%. Additionally, given that LLMs designed for code-specific tasks, such as CodeLlama, may struggle to generate personalities based on problem descriptions, we recommend using GPT-4o as the default LLM for personality generation, as demonstrated in our methodology.

}

\begin{table}[t]
\centering
\small
\renewcommand{\arraystretch}{1.3}
\begin{tabular}{l c c c}
\hline
\multicolumn{1}{l}{\textbf{LLM}} & \makecell[c]{\textbf{Direct}} & \makecell[c]{\textbf{Self}} & \makecell[c]{\textbf{GPT-4o}} \\
\hline
GPT-4o         & 78.3\%            & 84.3\%                    & -                 \\ 
GPT-4o mini    & 69.3\%            & 82.7\%                 & 82.2\%                 \\ 
Llama3.1      & 69.8\%            & 74.9\%                 & 81.0\%                 \\ 
Qwen-Long           & 68.4\%            & 74.5\%                 & 80.8\%                 \\ 
DeepSeek-Coder      & 74.9\%            & 84.8\%                 & 85.7\%                 \\ 
Codestral      & 64.2\%            & 70.7\%                 & 73.8\%                 \\ 
CodeLlama & 43.3\%            & -                     & 53.2\%                 \\ \hline
\end{tabular}
\captionsetup{singlelinecheck=off}
\centering
\caption{Comparison of pass rates between using each LLM individually for personality generation (Column ``Self'') and using GPT-4o consistently for personality generation}
\label{tab:SelfSota}
\end{table}

\section{Single Dimension Ablation Experiment}
\label{apendix:SingleDimension}

\begin{table}[t]
\centering
\small
\begin{tabular}{lccc}
\hline
\textbf{LLM} & \textbf{Direct} & \textbf{S/N Only} & \textbf{Personality} \\
\hline
GPT4o & 78.2\% & 78.5\% & 84.3\% \\
GPT4o mini & 69.3\% & 70.3\% & 82.2\% \\
llama 3.1 & 69.8\% & 72.2\% & 81.0\% \\
Qwen & 68.4\% & 74.2\% & 80.8\% \\
Deepseek & 74.9\% & 74.2\% & 85.7\% \\
Codestral & 64.2\% & 57.6\% & 73.8\% \\
Codellama & 43.3\% & 43.3\% & 46.8\% \\
\hline
\textbf{Avg. Change} & - & 0.3\% & 9.5\% \\
\hline
\end{tabular}
\caption{Ablation experiment results of considering only the sensing/intuition dimension}
\label{tab:SvsN}
\end{table}

\modified{

The distribution of MBTI types generated by GPT-4o for each dataset is primarily concentrated around two types: INTJ and ISTJ. Thus we conduct the ablation experiment to investigate whether considering only the sensing/intuition dimension suffice.

Using the representative MBPP Sanitized dataset, we prompted the LLM to adopt the role of a programmer with the specified S/N property for each task. The results, shown in the table below, reveal that using only the S/N property improves code generation accuracy for 4 out of 7 LLMs, with an average pass rate increase of 0.3\%. By comparison, prompts incorporating the full MBTI properties achieve a higher average pass rate improvement of 9.5\%. This result reveals the adequacy of using MBTI modeling in its entirety rather than solely one dimension.

}

\begin{table*}[t]
\centering
\small
\begin{tabularx}{\textwidth}{l|*{3}{>{\centering\arraybackslash}X}|*{3}{>{\centering\arraybackslash}X}}
\toprule
\textbf{LLM} & \multicolumn{3}{c}{\textbf{Direct}} & \multicolumn{3}{|c}{\textbf{Personality}} \\ 
\cmidrule(lr){2-7} 
 & \textbf{Pass@1} & \textbf{Pass@5} & \textbf{Pass@10} & \textbf{Pass@1} & \textbf{Pass@5} & \textbf{Pass@10} \\ 
\midrule
GPT-4o & 78.2\% & 80.1\% & 82.0\% & 84.3\% & 91.6\% & 93.9\% \\ 
GPT-4o mini & 69.3\% & 74.7\% & 77.8\% & 82.2\% & 86.4\% & 92.0\% \\ 
Llama3.1 & 69.8\% & 74.0\% & 76.6\% & 81.0\% & 85.5\% & 92.3\% \\ 
Coderstral & 64.2\% & 73.8\% & 79.2\% & 73.8\% & 85.0\% & 92.7\% \\ 
CodeLlama & 43.3\% & 46.6\% & 47.1\% & 46.8\% & 49.6\% & 51.2\% \\ 
\bottomrule
\end{tabularx}
\caption{Comparison of pass@k achieved by Direct and Personality methods on MBPP Sanitized dataset}
\label{tab:pass_k}
\end{table*}

\section{Pass@5 and Pass@10 Results}
\modified{

Following former code generation work\citep{chen2021evaluating, pass_at_1}, we add Pass@k as a complement to our evaluation metrics. Specifically, given a task, a LLM generates k programs. The task is solved if any generated programs pass all test cases. We compute the percentage of solved tasks in total requirements as Pass@k. The formula is:

$$\text{Pass@k} = \frac{1}{N} \sum_{i=1}^{N} 1*\left( \text{Correct}_i \in \text{Top-}k(\text{Choices}_i) \right)$$

We choose $k \in \{1, 5, 10\}$, and we select MBPP Sanitized dataset for this experiment. Since the APIs of Qwen-long and Deepseek Coder V2 do not support generating top k options simultaneously, we only measure the Pass@k of the rest five models. The results are listed in Table \ref{tab:pass_k}. The results of Pass@5 and Pass@10 are the same as the conclusions about MBPP Sanitized in RQ1, which help strengthen the robustness of the findings.

}

\section{Resource Consumption Report}

\begin{table}[t]
\centering
\small
\begin{tabular}{lccc}
\hline
\textbf{LLM} & \textbf{Unit Price} & \makecell[c]{\textbf{GPU Time}} & \textbf{Total Cost} \\
\hline
GPT4o & \$5 & - & \$1.75 \\
GPT4o mini & \$0.3 & - & \$0.105 \\
llama 3.1 & - & 11.5h & \$13.8 \\
Qwen & \$0.28 & - & \$0.098 \\
Deepseek & \$0.28 & - & \$0.098\\
Codestral & \$0.5 & - & \$0.175 \\
Codellama & - & 1.5h & \$1.8 \\
\hline
\end{tabular}
\caption{Resource consumption of different LLMs on one experiment round, including direct prompting, personality suggestion, and personality-guided code generation. The unit price is for 1 million token outputs, and the GPU time is based on A800, which is \$1.2 for one hour of calculation.}
\label{tab:Resource}
\end{table}

\modified{
Regarding computational costs, using the representative MBPP Sanitized dataset (427 tasks), one experiment round, including direct prompting, personality suggestion, and personality-guided code generation, consumes 0.35M tokens, costing about \$1.75 with GPT-4o (\$0.004 per task). The costs are significantly lower for other LLMs. The resource consumption of different models is listed in Table \ref{tab:Resource}. The results reveals that the overall cost of our method is rather efficient. For llama 3.1 and codellama are deployed on a computing cluster with 4 A800 GPUs, thus the cost is more than API calls. 
}

\end{document}